\documentclass[twoside,11pt]{article}
\usepackage{epsfig}

\newcommand{\be}{\begin{eqnarray}}

\newcommand{\beq}{\begin{equation}}
\newcommand{\eeq}{\end{equation}}

\newcommand{\ee}{\end{eqnarray}}
\newcommand{\ba}{\begin{array}}
\newcommand{\ea}{\end{array}}

\begin{document}
\thispagestyle{empty}
\title{\bf Extraction of radiative decay width for the non-strange
partner of $\Theta^+$}
\author{Ya. Azimov$^{1}$, V. Kuznetsov$^2$, M. V. Polyakov$^{1,3}$,
I. Strakovsky$^4$\\
\\
$^1$St. Petersburg Nuclear Physics Institute,\\ Gatchina, 188300 Russia \\
$^2$ Institute for Nuclear Research,\\ Moscow, 117312 Russia\\
$^3$ Institut f\"ur Theor. Physik -II, Ruhr-Universit\"at,\\ D-44780
Bochum, Germany\\
$^4$ Center for Nuclear Studies, Physics Department, \\ The George
Washington University,\\ Washington, DC 23606 USA} \maketitle

\begin{abstract}
Using the results of the GRAAL collaboration on the $\eta$
photoproduction from the neutron target, we attempt to extract the
partial radiative width of the possible new nucleon resonance
$N^*(1675)$. The obtained estimates support this resonance to be a
very attractive candidate for the non-strange member of the exotic
antidecuplet of baryons -- a partner of the $\Theta^+$ pentaquark.
Our phenomenological value for the transition magnetic moment
$\mu(n^*\,n)$, appears to be in good agreement with predictions of
the Chiral Quark Soliton Model.

\end{abstract}


\vspace{0.2cm} The first independent evidences for the exotic
baryon $\Theta^+$, with strangeness $+1$, obtained in reactions
$\gamma{}^{12}{\rm C}$~\cite{Osaka} and $K^+{\rm Xe}$~\cite{ITEP},
were followed by important confirmations in more than ten
experiments  (see recent experimental reviews~\cite{Hicks}). We
still know rather little about properties of the exotic
$\Theta^+$, even its very existence has not yet been firmly
established, see discussion of the null experimental results in
ref.\cite{Hicks}.

However, if the $\Theta^+$ does really exist, we have to deal from
now on with a new type of baryonic multiplets in terms of the
flavour $SU(3)$ group, most probably, the antidecuplet. This
corresponds to predictions of the Chiral Quark Soliton Model
(ChQSM)~\cite{DPP97} (see also recent theoretical
reviews~\cite{pras-kop} and detailed references therein). In terms
of quarks, the antidecuplet cannot be composed of three quarks,
contrary to all previously known baryons. Its minimal quark
configuration should be the 5-quark one (4 quarks and 1
antiquark). If so, it would be of great importance, in parallel to
further studies of the $\Theta^+$, to find and investigate other
members of the antidecuplet. Also very interesting would be to
understand internal structure of the new baryons in terms of
quarks.

Recently the GRAAL collaboration \cite{Kuznetsov} provided an indication for a nucleon
resonance  $N^*(1675)$ in the process $\gamma n\to \eta n$. Its mass and seemingly
narrow width correspond to expected properties of the non-strange partner of the narrow
$\Theta^+$, see~\cite{AAPSW,DP03,GP05}. The fact that this nucleon state is excited by
the photon preferably on the neutron target (not on the proton one) also favours its
antidecuplet nature~\cite{PR03}.

In this note we try to estimate the partial radiative width
$\Gamma(n^*\to n\gamma)$ using the results of the GRAAL
collaboration \cite{Kuznetsov,KuznetsovTrento}. It is directly
related to the corresponding transition multipole moment (magnetic
dipole, according to ChQSM expectations). Though such a moment may
be calculated theoretically in various approaches (ChQSM, sum
rules, and so on), it is especially interesting for the quark
description, where this moment is very sensitive to the hadron
internal structure.

The refs.~\cite{Kuznetsov,KuznetsovTrento} reveal an irregular
behaviour of the cross section for the process $\gamma n\to \eta
n$ near the invariant mass  $W=1675$~MeV. In
ref.~\cite{KuznetsovTrento}, the differential cross section at
scattering angle around $140^o$ has been fitted by a smooth
background plus a Gaussian peak. Assuming, for simplicity,
incoherence of the resonance and background, the peak contribution
has been fitted to the form:
\be \frac{d \sigma_{\rm visible\
peak}}{d\Omega}(W) = 0.19\ e^{-{(W-M_R)^2/(\Delta W)^2}}\, \mu{\rm
b}\cdot{\rm sr}^{-1}\, , \label{fit} \ee
where $W$ is the invariant mass of the $\eta n$ (and/or $\gamma n$)
system, $M_R=1675$~MeV and $\Delta W=30$~MeV. The measured peak mass
agrees with the candidate value $M=1680$~MeV for the assumed partner
of $\Theta^+(1540)$, as extracted from the Partial-Wave Analysis of
$\pi N$ scattering~\cite{AAPSW}. The visible peak width is
consistent with the expected smearing effects due to the Fermi
motion and experimental resolution. Therefore, assuming that the
peak corresponds to a nucleon resonance, we conclude that its true
width should be considerably smaller. Then, we can write for the
integrated cross section: \be \int\ dW\ \frac{d \sigma_{\rm
res}}{d\Omega}(W) =\frac{\pi}{4 k^2_{\gamma}}\,
 \frac{\Gamma_{\gamma n}\Gamma_{\eta n}}{\Gamma_{\rm tot}}\,.
\label{eq1} \ee
Here $\Gamma_{\rm tot}$ is the true total width of the resonance,
while $\Gamma_{\gamma n}$ and $\Gamma_{\eta n}$ are its partial
widths for the decay modes to ${\gamma n}$ and ${\eta n}$
respectively; $k_{\gamma}=574$~MeV is the c.m. relative momentum
for the initial $\gamma n$ state at the resonance energy. We have
also assumed that the nucleon resonance under discussion is of
spin 1/2 (as suggested by ChQSM for the flavour partners of the
$\Theta^+$), so its cross section has a flat angular dependence.
The area of integration over the invariant mass in Eq.~(\ref{eq1}),
which is effectively $\sim\Delta W$, should overlap the experimental
resolution and Fermi motion smearing. It is assumed to be larger
than $\Gamma_{\rm tot}\,.$ Such assumption is consistent with the
Monte Carlo simulation (see Fig.3 of ref.\cite{Kuznetsov}), which
suggests $\Gamma_{\rm tot}\approx10$~MeV. This agrees with the
earlier theoretical estimate $\Gamma_{\rm tot}\leq
10$~MeV~\cite{AAPSW}.

Now, using Eq.~(\ref{eq1}) and taking the l.h.s. integral of the
visible peak cross section (\ref{fit}), we obtain expression for
the radiative partial width of $n^*$: \be \Gamma\left(n^*\to \gamma
n\right)= \frac{4 k^2_{\gamma}\, \Delta W}{\sqrt{\pi}\,{\rm Br(\eta
n)}}\,  \left.  \frac{d\sigma_{\rm visible\
peak}}{d\Omega}\right|_{W=M_R} \approx \frac{1.08\cdot 10^{-2}}{{\rm
Br(\eta n)}} \ {\rm MeV}.  \ee The radiative partial width is simply
related with the transition magnetic moment $\mu(n^*\to n)$ : \be
\Gamma(n^*\to \gamma n)=\left(\frac{\mu}{\mu_N}\right)^2\
\frac{\alpha}{M_N^2}\ k_\gamma^3\, , \ee where $\mu_N$ is the nuclear
magneton, $\alpha=1/137$, and $k_\gamma$ is the photon momentum in the
rest frame of the decaying resonance ($i.e.$, the same as above). Now,
at last, we arrive at an estimate: \be |\mu(n^*\to n)|=
\frac{0.083}{ \sqrt{{\rm Br(\eta n)}}}\,\mu_N \,. \label{result}
\ee Surely, this estimate is rather rough, but corresponds to the
present quality of experimental data. In particular, the data do
not allow to separate different partial-wave contributions. Therefore,
we do not model in detail the non-resonant background for the $N^*$,
and even have not explicitly accounted for a possible effect of nearby
wide resonances (according to RPP~\cite{PDG04}, they could be
$S_{11}(1650),$ $D_{15}(1675),$ $F_{15}(1680),$ $D_{13}(1700),$
$P_{11}(1710),$ and $P_{13}(1720)$, all with widths $\sim
100-200$~MeV).  Moreover, we have assumed in our estimates that the
peak seen by the GRAAL collaboration corresponds to a narrow resonance
with definite quantum numbers $J^P= 1/2^+$, as is expected in the
ChQSM (the case of $1/2^-$ would look similar, but with $\mu(n^*\to n)$
being the transition electric dipole moment).  Nevertheless, we believe
that our estimate (\ref{result}) may be reasonable and close to
reality, say, within a factor of two.

Eq.~(\ref{result}) still depends on the branching ratio of the decay $n^*
\to \eta n$. For its value we employ estimates obtained in the framework
of ChQSM~\cite{DPP97,AAPSW,Michal}, as well as the phenomenological analysis
of ref.~\cite{GP05}. It is expected that the possible range of the $\eta n$
branching is $0.05 - 0.4$.  Using this range, we obtain, as the final result,
that the transition magnetic moment $\mu(n^*\to n)$ lies in the interval
$$|\mu(n^*\to n)|=(0.13 -0.37)\, \mu_N\,,$$ which corresponds to
$\Gamma_{\gamma n}=(27 - 216)$~keV.

Even keeping in mind all uncertainties of the above theoretical
estimates, of the present phenomenological analysis, and of
experimental errors in the fit (\ref{fit}), the obtained numbers
are self-consistent and agree quite well with ChQSM expectations
for this transition magnetic moment to be $(0.10 -
0.56)\,\mu_N$~\cite{GKPP05}. It is important also to note that the
extracted moment is numerically small (for comparison,
$\mu(\Delta\to N)\approx 3$). This qualitative feature is in
agreement with the prediction of ChQSM that in the
non-relativistic limit the corresponding transition moment
vanishes~\cite{PR03}.

This smallness correlates with the ChQSM expectation of an
anomalously small width of the $\Theta^+$~\cite{DPP97}; the reason
for smallness, in both cases, is the formal cancellation of
various coupling constants, which becomes exact in the
non-relativistic limit. It was suggested years ago~\cite{az70}
that since hadrons are systems with an internal structure, some of
them may have the structure essentially different from that of
conventional hadrons. Such structure inconsistency could suppress
their couplings with the conventional hadrons, leading to small
production cross sections and decay widths, despite the strong
interaction nature of the processes. A recent calculation of the
$\Theta^+$ width~\cite{DP05}, in terms of quark configurations in
effective mean field, seems to explain its smallness by the
essential difference between quark structures for the initial
(mainly 5-quark $\Theta$) and final (mainly 3-quark nucleon)
baryons, thus supporting the assumed possibility.

Let us emphasize once more the absence of the corresponding resonance-like structure in
the $\gamma p \to \eta p$ process \cite{Kuznetsov,KuznetsovTrento}.
It implies that the ratio of magnetic transitions
$\mu(p^*p)/\mu(n^*n)$ should be less than $\sim 1/3$, otherwise the corresponding
peak would be visible in data on $\eta$ photoproduction off the proton target, having
higher statistics and better quality. This supports the (mainly) antidecuplet nature of
the discussed resonance.

To summarize, we have tried to extract the radiative decay
parameters for a new nucleon resonance which could be a partner of
the $\Theta^+$. Even with existing preliminary data, our results
seem to be self-consistent in the framework of the Chiral Quark
Soliton model.

\vspace{0.2cm} {\small The work was partly supported by the
Russian State Grant RSGSS-1124.2003.2, by the Russian-German
Collaboration Treaty (RFBR, DFG), by the COSY-Project J\"ulich, by
the Sofja Kowalewskaja Programme of Alexander von Humboldt
Foundation, BMBF, DFG, by the U.~S.~Department of Energy Grant
DE--FG02--99ER41110, by the Jefferson Laboratory, and by the
Southeastern Universities Research Association under DOE Contract
DE--AC05--84ER40150.}

\end{document}